\documentclass[twocolumn,dvipsnames]{aastex62}
\pdfoutput=1 
\usepackage{amsmath,amstext}
\usepackage{apjfonts}
\usepackage[figure,figure*]{hypcap}
\usepackage{ulem}
\usepackage{xspace}

\usepackage{xfrac}
\usepackage{graphicx}
\usepackage{natbib}
\usepackage{amssymb}
\usepackage{amsmath}
\usepackage{array,booktabs}
\usepackage{mathptmx}
\usepackage{booktabs}
\usepackage{hyperref}
\usepackage{float}
\usepackage{todonotes}

\DeclareSymbolFont{cmletters}{OML}{cmm}{m}{it}
\DeclareMathSymbol{v}{\mathalpha}{cmletters}{"76}

\newcommand{\msun}{\,{M_{\odot}}}
\newcommand{\cm}{\,{\rm cm}}
\newcommand{\s}{\,{\rm s}}	
\newcommand{\erg}{\,{\rm erg}}

\newcommand{\Mej}{\,{M_{\rm ej}}}

\newcommand{\Ah}{\,{a8\delta0}}
\newcommand{\Am}{\,{a2\delta0}}
\newcommand{\Al}{\,{a1\delta0}}
\newcommand{\C}{\,{a1\delta1}}

\newcommand{\appropto}{\mathrel{\vcenter{
			\offinterlineskip\halign{\hfil$##$\cr
				\propto\cr\noalign{\kern2pt}\sim\cr\noalign{\kern-2pt}}}}}
\usepackage{xcolor}

\shorttitle{Collapsar Black Holes are Likely Born Slowly Spinning}
\shortauthors{Gottlieb, Jacquemin-Ide, Lowell \& Tchekhovskoy}

\begin{document}
\title{Collapsar Black Holes are Likely Born Slowly Spinning}
	\author[0000-0003-3115-2456]{Ore Gottlieb}
	\email{ore@northwestern.edu}
	\affiliation{Center for Interdisciplinary Exploration \& Research in Astrophysics (CIERA), Physics \& Astronomy, Northwestern University, Evanston, IL 60202, USA}

	\author[0000-0003-2982-0005]{Jonatan Jacquemin-Ide}
	\affiliation{Center for Interdisciplinary Exploration \& Research in Astrophysics (CIERA), Physics \& Astronomy, Northwestern University, Evanston, IL 60202, USA}

	\author[0000-0002-2875-4934]{Beverly Lowell}
	\affiliation{Center for Interdisciplinary Exploration \& Research in Astrophysics (CIERA), Physics \& Astronomy, Northwestern University, Evanston, IL 60202, USA}

	\author[0000-0002-9182-2047]{Alexander Tchekhovskoy}
	\affiliation{Center for Interdisciplinary Exploration \& Research in Astrophysics (CIERA), Physics \& Astronomy, Northwestern University, Evanston, IL 60202, USA}

	\author[0000-0002-9182-2047]{Enrico Ramirez-Ruiz}
	\affiliation{Department of Astronomy and Astrophysics, University of California, Santa Cruz, CA 95064, USA}
 
\begin{abstract}

    Collapsing stars constitute the main black hole (BH) formation channel, and are occasionally associated with the launch of relativistic jets that power $ \gamma $-ray bursts (GRBs). Thus, collapsars offer an opportunity to infer the natal (before spin-up/down by accretion) BH spin directly from observations. We show that once the BH saturates with large-scale magnetic flux, the jet power is dictated by the BH spin and mass accretion rate. Core-collapse simulations by \citet{Halevi2023} and GRB observations favor stellar density profiles that yield an accretion rate $ \dot{m} \approx 10^{-2} \msun~\s^{-1} $, weakly dependent on time. This leaves the spin as the main factor that governs the jet power. By comparing the jet power to characteristic GRB luminosities, we find that the majority of BHs associated with jets are likely born slowly spinning with a dimensionless spin $ a \simeq 0.2 $, or $ a \lesssim 0.5 $ for wobbling jets, with the main uncertainty originating in the unknown $ \gamma $-ray radiative efficiency. This result could be applied to the entire core-collapse BH population, unless an anti-correlation between the stellar magnetic field and angular momentum is present. In a companion paper, \citet{Jacquemin2023}, we show that regardless of the natal spin, the extraction of BH rotational energy leads to spin-down to $ a \lesssim 0.2 $, consistent with gravitational-wave observations. We verify our results by performing the first 3D general relativistic magnetohydrodynamic simulations of collapsar jets with characteristic GRB energies, powered by slowly spinning BHs. We find that jets of typical GRB power struggle to escape from the star, providing the first numerical indication that many jets fail to generate a GRB.
		
\end{abstract}
	
	\section{Introduction}\label{sec:introduction}
    When the core of a massive star exhausts its nuclear fuel, it collapses under its own gravity to form a proto-neutron star (PNS). If the PNS accretes mass above $ M_{\rm NS,max} \gtrsim 2.2\msun $ \citep{Margalit2017,Aloy2021,Obergaulinger2022}, then it undergoes a further collapse to form a black hole (BH). As a non-negligible fraction of all core-collapse supernovae ultimately produce a BH \citep[e.g.,][]{Kochanek2015}, collapsing stars constitute the main BH formation channel in the Universe. The BH initially interacts with the dense stellar envelope, gains mass and angular momentum. At these early times, the BH vicinity is opaque to electromagnetic (EM) radiation, so the BH early evolution is observationally out of reach.
    
    Long after the BH formed and the progenitor star is gone, the BH may become involved in new astrophysical processes, such as accretion or merger, that trigger EM emission from which we can infer the BH properties, e.g. mass and spin \citep[see][for reviews]{Middleton2016,Reynolds2021}.
    In addition to EM signals, the first gravitational wave detections of binary BH mergers by LIGO/Virgo/KAGRA (LVK) also provide clues regarding the BH nature. For example, different studies have consistently found that pre-merger BHs spin slowly \citep[][but see also \citealt{Safarzadeh2020}]{Farr2017,Tiwari2018,Roulet2019,Abbott2020,Hoy2022}.
    However, as we show in a companion paper, the inferred low spin might be an indication of a substantial BH spin-down rather than a low natal spin -- the spin with which the BH forms \citep{Jacquemin2023}. Thus, while EM and gravitational wave detections can shed light on the resultant BH properties after mass accretion, and spin-down/up, the natal BH properties are challenging to study observationally. It thus remains unclear whether BHs are born slowly spinning or are spun down after their formation.

    The natal BH spin can be estimated on theoretical grounds.
    \citet{Fuller2019,Belczynski2020} found efficient angular momentum transport in stars via magnetic Tayler instability, such that newly born BHs in massive stars maintain a low dimensionless spin $ a \approx 10^{-2} $, where $ a $ varies from $ a = 0$ (non-spinning BH) to $ a = 1 $ (maximally-spinning BH). This result is consistent with the natal BH spin that follows the collapse of a millisecond PNS
    \begin{equation}\label{eq:aNS}
    a \simeq 0.03\left(\frac{R_{\rm NS}}{10~{\rm km}}\right)^2\frac{1\rm ms}{P}\frac{3\msun}{M_{\rm NS,max}}~,
    \end{equation}
    where $ R_{\rm NS} $ is the PNS radius, and $ P $ is the PNS spin period. We stress that both of those quantities could be significantly different than the canonical values in Eq.~\eqref{eq:aNS}, depending on the equation of state, stellar angular momentum, etc. In such cases, the PNS may give birth to a rapidly spinning BH \citep[e.g.,][]{Aloy2021,Obergaulinger2022}.
    Additionally, there is a growing evidence that the majority of BHs' very massive progenitors belong to multi-star systems  \citep{Kobulnicky2007,Smith2011,Sana2012,Duchene2013,Offner2014}. The companion star may spin up the BH progenitor star, and consequently the newly born BH itself.
    However, stellar population simulations show that the rotational velocity of the progenitor external shells is just a few hundred $ {\rm km~s^{-1}} $ \citep{deMink2013}. For a strong coupling between the stellar core and envelope, as suggested by \citet{Fuller2019}, BHs in binary stars would also form with $ a \ll 1 $ \citep{Tchekhovskoy2015}.
    
    Massive stars in binary systems are considered to be the progenitors of long $ \gamma $-ray bursts \citep[GRBs;][]{DeBonder1998,Fryer2007,Yoon2010}, as those can supply enough angular momentum to the star to form the accretion disk required for launching relativistic jets \citep[e.g.,][]{Izzard2004,Fryer2005,Petrovic2005,Lee2006,Woosley2006,Cantiello2007}. The relativistic jets that generate the GRB are considered to be electromagnetically driven \citep[e.g.,][]{Lyutikov2003,Kawanaka2013}, extracting rotational energy from the BH \citep[][hereafter BZ]{Penrose1971,Blandford1977}.
    Therefore, GRB jets can carry the information from the BH and provide a direct connection between the BH spin and the observed GRB power. This enables us to overcome the above uncertainties in stellar evolution models, and constrain the natal BH properties from observations, through their imprinted properties on the expelled outflows.
    
    In this \emph{Letter}, we constrain the natal BH spin from observations for the first time. Many previous studies have focused on rapidly spinning BHs as the central engines powering relativistic GRB jets \citep{Komissarov2009,Janiuk2010,Tchekhovskoy2011,Aloy2021,Bavera2022,Gottlieb2022a,Gottlieb2022c,Fujibayashi2022}. Here, we argue analytically (\S\ref{sec:theory}) and numerically (\S\ref{sec:numerics}) that GRB observables suggest that the majority of BHs form with a low spin. This makes low BH spins more attractive for GRB jet launching than previously thought. We verify our conclusions by presenting the first 3D general-relativistic magnetohydrodynamic (GRMHD) simulations of collapsar jets from slowly spinning BHs. We discuss the implications for jets and their progenitor stars in \S\ref{sec:conclusions}.
 
 \section{Why are jet-associated Black Holes slowly spinning?}\label{sec:theory}

    \citet{Gottlieb2022a} performed the first 3D GRMHD simulations of collapsars that follow BZ-powered relativistic jets from the BH to outside of the star. They fixed the BH dimensionless spin to $ a = 0.8 $, and explored the effect of the magnetic field strength on relativistic jet launching. The simulations showed that if the magnetic field amplitude is below a certain threshold, which depends on the stellar core density, then the jets fail to launch. Interestingly, \citet{Gottlieb2022a,Gottlieb2022c} found that even when the magnetic field is set at the minimum threshold for jet launching, the emerging jet is orders of magnitudes more powerful than the observed characteristic GRB luminosity\footnote{If the inner stellar mass density profile is roughly flat, $ \rho \propto r^0 $, the density needs to be low in order to result in a typical stellar mass, $ \sim 20~\msun $: this would result in a jet of typical GRB energy. However, such a density profile introduces two problems: (i) jet power shows significant time evolution that is in tension with observations, and (ii) jet engine life time that is shorter than that of observed long GRBs \citep{Gottlieb2022a}.}. This begs the question of what pre-launching conditions are required for powering typical GRBs.
    
    \citet{Gottlieb2022a} interpreted the above results with formulating a jet launching criterion that requires the BZ power, which scales quadratically with the BH spin and the magnetic flux threading the BH, to overcome the accretion power of the infalling gas. However, since they did not consider different BH spins, their jet launching and power were solely dictated by the initial magnetic field strength and accretion rate, so they could not have distinguished between the role of the magnetic field and the BH spin.
    \citet{Komissarov2009} proposed that jets are launched successfully so long as their Alfv\'{e}n velocity is higher than the free-fall velocity of the inflowing gas. This is to allow the magnetohydrodynamic waves to escape from the BH ergosphere and constitute the emerging jet. According to this criterion, it is a strong enough magnetic flux, rather than a high enough BZ power, that enables a BH to launch jets against the onslaught of the infalling stellar envelope\footnote{\citet{Komissarov2009} found a weak dependency of the jet launching on the BH spin, denoted as $ \kappa $ in their work. As $ \kappa $ was found to change by up to a factor of 2 with the BH spin, its variance is negligible compared to the change by order of magnitude of the magnetic field threading the BH.}. Furthermore, jet launching is numerically found to be sustained once the disk becomes magnetically arrested (MAD), which in turn takes place when the dimensionless magnetic flux (normalized by the mass accretion rate) reaches a certain threshold, $ \phi_{\rm H} \approx 50 $ \citep[e.g.,][]{Tchekhovskoy2015b}. In conclusion, the jet launching criterion depends solely on the magnetic flux, and is independent of the BH spin. In contrast, the jet power solely depends on the BH spin and mass accretion rate, while the magnetic flux is saturated at the MAD value and thus has no freedom to control the jet power.
    Our numerical simulations support these conclusions (\S\ref{sec:numerics}). 
    
    Given the need for a sufficiently strong magnetic field to launch GRB jets inside a collapsing star, one way of achieving the desired (lower) characteristic power of GRB jets is by considering a lower BH dimensionless spin. The electromagnetic jet luminosity,
    \begin{equation}\label{eq:L}
        L_j = \eta(a)\dot{m}c^2~,
    \end{equation}
    is defined as the electromagnetic energy flux leaving the BH,
    where $ \dot{m} $ is the mass accretion rate onto the BH. The jet launching efficiency solely depends on the dimensionless BH spin \citep{Lowell2023},
    \begin{equation}\label{eq:eta}
        \eta(a) = 1.063a^4+0.395a^2~.
    \end{equation}
    Equations \eqref{eq:L}, \eqref{eq:eta} demonstrate that if a jet is launched in a MAD state (in which the BH magnetic flux is saturated at the maximum, MAD, value), then its luminosity is governed only by the accretion rate and the BH spin.
    
    The accretion rate is determined by the mass density profile of the stellar envelope. Recently, \citet{Halevi2023} examined the stellar density profile evolution during the PNS stage, between the onset of the core-collapse and the formation of the BH. They found that all stellar evolution models, which feature steep density profiles with a power-law index of $ \alpha = 2.5 $ at the onset of the collapse, consistently converge to a shallower density profile with $ \alpha = 1.5 $ at the BH formation time. In a free-fall collapse, a power-law density profile leads to mass accretion rate \citep{Gottlieb2022a}
    \begin{equation}\label{eq:mdot}
        \dot{m}(\alpha) \sim t^{1-2\alpha/3}~.
    \end{equation}
    Namely, $ \dot{m}(\alpha=1.5) \sim {\rm const} $. For a roughly constant jet launching efficiency, as expected in a saturated MAD state \citep[e.g.,][and \S\ref{sec:numerics}]{Tchekhovskoy2015}, the jet power remains constant as well, as implied by GRB observations \citep{Mcbreen2002}.
    
    If indeed $ \alpha = 1.5 $ at the onset of the BH formation is a universal value, one can derive a universal accretion rate. All mass density profiles obtained by \citet{Halevi2023} roughly coincide with the initial profile used in the simulations performed by \citet{Gottlieb2022c}. In these models, \citet{Gottlieb2022c} used a representative total stellar mass (including the BH mass) of $ 18.2~\msun $. Their numerical results featured a roughly constant accretion rate of $ \dot{m} \approx 5\times 10^{-3} \msun~\s^{-1} $, but with a moderate decrease owing to suppression of accretion from the outflows. This accretion rate might increase linearly with the stellar mass, which may be larger by a factor of a few\footnote{The accretion rate also increases with the BH mass, $ \dot{m} \sim v_{\rm ff} \sim M^{1-\alpha/3} $, where $ v_{\rm ff} $ is the free-fall velocity, and thus $ \dot{m} \sim M^{0.5} $ for $ \alpha = 1.5 $. For $ \dot{m} \approx 10^{-2} \msun~\s^{-1} $, the BH mass may double its mass during the jet launching, corresponding to a small non-linear correction to the mass accretion rate.}. Thus, for a canonical accretion rate $ \dot{m} \approx 10^{-2} \msun~\s^{-1} $ \citep[as was also found in core-collapse simulations, e.g.,][]{Obergaulinger2022}, Eqs. \eqref{eq:L} and \eqref{eq:eta} dictate that the jet luminosity depends on the BH spin,
    \begin{equation}\label{eq:L2}
        L_j \approx 2\times 10^{52}(1.063a^4+0.395a^2)\left(\frac{\dot{m}}{10^{-2}~\msun~\s^{-1}}\right)~\erg~\s^{-1}~.
    \end{equation}
    
    For a characteristic observed isotropic equivalent $ \gamma $-ray luminosity of $ L_{\gamma,{\rm iso}} \simeq 3\times 10^{52}~\erg~\s^{-1} $ \citep{Wanderman2010}, and opening angle $ \theta_j \simeq 0.1 $ \citep{Goldstein2016}, the intrinsic two-jet $ \gamma $-ray luminosity is $ L_\gamma \equiv L_{\gamma,{\rm iso}}\left[1-{\rm cos}\left(\theta_j\right)\right] \simeq 1.5\times 10^{50}~\erg~\s^{-1} $.
     The jet power is determined by the poorly constrained $ \gamma $-ray radiative efficiency $ \epsilon_\gamma $ \citep{Eichler2005}, whose wide range from $ \epsilon_\gamma \ll 1 $ \citep{Frail2001} to $ \epsilon_\gamma \approx 0.8 $ \citep[e.g.,][]{Panaitescu2002} introduces a significant uncertainty. We choose a fiducial value of $ \epsilon_\gamma = 0.5 $, so the total jet luminosity is $ L_j = L_\gamma/\epsilon \simeq 3\times 10^{50}(0.5/\epsilon_\gamma)(\theta_j/0.1)^2 ~\erg~\s^{-1} $. Plugging in Eq. \eqref{eq:L2}, the corresponding BH dimensionless spin is
    \begin{equation}
    a \simeq 0.18\left(\frac{0.5}{\epsilon_\gamma}\right)^{0.5}\left(\frac{\theta_j}{0.1}\right)~,
    \end{equation}
    where the dependency on $ \epsilon_\gamma, \theta_j $, assumes $ \eta \sim a^2 $, which is valid at low spins, and is a rough estimate to within a factor of a few at high spins.

    In addition to the radiative efficiency, another difficulty in the above analysis is the non-trivial conversion of isotropic equivalent energy to total jet energy. \citet{Gottlieb2022c} showed that jets inevitably exhibit a wobbling motion, as we also find in all of our numerical models (\S\ref{sec:numerics}). The jet wobbling is caused by the spontaneous tilt of the disk, induced by the stochastic torques applied to the disk by infalling gas. Consequently, a given observer will only see a fraction of the jet energy, so that the \emph{observed} jet energy is in fact an order of magnitude lower than the true jet energy \citep{Gottlieb2022e}. In such cases, the total GRB jet energy is $ L_j \simeq 3\times 10^{51}(0.5/\epsilon_\gamma)(\theta_w/0.3)^2 ~\erg~\s^{-1} $, where $ \theta_w $ is the wobble amplitude. Eq.~\eqref{eq:L2} dictates that in this case $ a \simeq 0.48 $. However, if the jet wobbles, the inferred radiative efficiency is likely lower than the real value, so that $ a \lesssim 0.48 $. \citet{Lowell2023} argued that cooled disk would have a milder equilibrium spin, $a = 0.3$, compared to non-cooled disk in which the equilibrium spin is $a=0.1$. Thus, the jet wobbling motion might probe the cooling state of collapsar accretion disks.

    \section{Numerical simulations}\label{sec:numerics}
    We investigate the emergence of GRB jets from slowly spinning BHs by performing the first collapsar simulations with a low BH spin, from jet launching to breakout.
    We build on the collapsar simulations that were recently carried out by \citet{Gottlieb2022c} using the code \textsc{h-amr} \citep{Liska2022}. The main difference between the initial physical setup of the models is the lower BH spin compared to \citet{Gottlieb2022c} who set it to be $ a = 0.8 $. The full setups are given below. We also conduct one simulation with an anisotropic mass distribution of the stellar envelope to study the effect of a low density polar region on the ability of weaker jets to break out from the star.
    Such anisotropic density distribution can emerge e.g. if the polar axis is vacated by rapid rotation of the envelope \citep[e.g.,][]{Fujibayashi2020}. However, stellar population simulations show that the stellar angular density profile is homogeneous to within a few percent, disfavoring a significant anisotropy \citep{deMink2013}. Nevertheless, the anisotropy may emerge in the post-collapse stage due to neutrino-antineutrino annihilation \citep{Eichler1989,Popham1999,MacFadyen1999}, magnetic outflows, or PNS-powered pre-cursor jets \citep[e.g.,][]{Burrows2007}. Those scenarios may form a low density funnel that mitigates the relativistic jet propagation through the envelope.
    The remaining initial conditions are identical between all simulations, and are summarized below.

    \subsection{Setup}

    We initialize the simulations with a central BH mass $ M = 4.2 \msun $ embedded in a stellar envelope of mass $ M_\star = 14\msun $ and radius $ R_\star = 4\times 10^{10} {\cm} $.
    The magnetic vector potential in the star has only an azimuthal component that scales as $ A_{\hat{\phi}}(r,\theta) \sim {\rm sin}(\theta)/r $, and drops to zero at the stellar surface. For code stability purposes, we set a maximum jet magnetization $ \sigma = 150 $, which is also the jet magnetization upon launching $ \sigma_0 $, or the jet asymptotic Lorentz factor. The stellar envelope {undergoes solid body rotation well below the centrifugal value throughout the star, with the specific angular momentum given by}\footnote{Note that this expression is a typo correction to the one in \citet{Gottlieb2022a,Gottlieb2022b}.}
    \begin{equation}
    l(r,\theta) =
    \begin{cases}
        \omega_0r^2{\rm sin}^2\theta & \text{for}\ r \le 70r_g,\\
        &\\ 
        \omega_0 (70r_g)^2{\rm sin}^2\theta &  \text{for}\ r > 70r_g, 
    \end{cases}
    \end{equation}
    where $ r_g \equiv GM/c^2 = 6.3\times 10^5~\cm $ is the BH gravitational radius, $\omega_0 = 50~{\rm s^{-1}} $.
    {We note that prior to the BH formation, \citet{Halevi2023} found that the innermost ($ \approx 10^{8.5} $ cm) stellar shells accelerate to high velocities. Since our simulations do not have self-gravity, we set the radial velocity to zero. However, those shells will accelerate as they free-fall and reach similar velocities. Importantly, the free-fall time of those shells is $\lesssim 0.1$~s, implying that after that time the progenitor structure is consistent with that of \citet{Halevi2023}.} {We have verified via a direct simulation that an identical setup with the initial radial velocity set to free-fall velocity converges to the same mass accretion rate after a short initial transient of $t \lesssim 0.1$~s.}
    
    We explore both isotropic and anisotropic mass density profiles, with the degree of anisotropy controlled by $ \delta $:
     \begin{equation}\label{eq:rho}
         \rho(r,\theta) = \rho_0 \left(\frac{r}{R_\star}\right)^{-1.5}\left(1 - \frac{r}{R_\star}\right)^{3}~{\rm sin}^\delta(\theta)~,
     \end{equation}
    where $ \rho_0 $ is set by $ M_\star $, and depends on the value of $ \delta $. We compare our simulations with the simulation of a rapidly spinning BH and $ \sigma_0 = 200 $ from \citet{Gottlieb2022c}. The parameters of the models are listed in Table~\ref{tab:models}.

    The numerical integration is performed on a spherical grid using an ideal gas law equation of state with index of $ 4/3 $. The radial grid is logarithmic from $ 0.9r_g $ to $ 10^6 r_g $. The number of cells in the base grid is $ 384 \times 96 \times 192 $ in the radial, polar and azimuthal directions, respectively. We use a local adaptive time-step and 3 levels of adaptive mesh refinement, tuned to provide approximately the same transverse resolution across the jets at all distances \citep[][]{Gottlieb2022c}. Thus, the maxim{um effective resolution of} the grid is $ 3072 \times 768 \times 1536 $. We tilt the {initial conditions and the} metric by $ 90^\circ $ to avoid numerical artifacts on the jet axis, which could emerge due to the discontinuity on the polar axis \citep{Gottlieb2022a}.

    \begin{table}
    	\setlength{\tabcolsep}{4.5pt}
    	\centering
    	\renewcommand{\arraystretch}{1.1}
    	\begin{tabular}{ c | c | c | c | c | c | c }
    		
    	Model & $ a $ & $ \delta $ & $ t_b~[\s] $ & $ t_s~[\s] $ & $ \Mej/M_\star~(9.2~\s) $ & $ \Mej/M_\star~(t_s) $ \\	\hline
            $ \Ah $ & $ 0.8 $ & $ 0 $ & $ 2.1 $ & $ 9.2 $ & 65\% \\ 
            $ \Am $ & $ 0.2 $ & $ 0 $ & $ 11.5 $ &  $ 15.5 $ & 52\% & 75\% \\ 
            $ \Al $ & $ 0.1 $ & $ 0 $ & $ 16 $ & $ 22 $ & 32\% & 79\% \\ 
            $ \C $ & $ 0.1 $ & $ 1 $ & $ 3.5 $ & $ 9.2 $ & 21\% \\ 
    	\end{tabular}
    
    \caption{The parameters of the different models. $ a $ is the BH dimensionless spin, $ \delta $ is the mass distribution anisotropic component power-law index, $ t_b $ is the shock breakout time from the star, $ t_s $ is the simulation duration, and $ \Mej/M_\star $ is the fraction of the stellar envelope that is unbound (hydrodynamically and magnetically).}
     
    \label{tab:models}
    \end{table}

    \subsection{Accretion \& Launching}\label{sec:accretion}
    
\begin{figure}
        \includegraphics[scale=0.33]{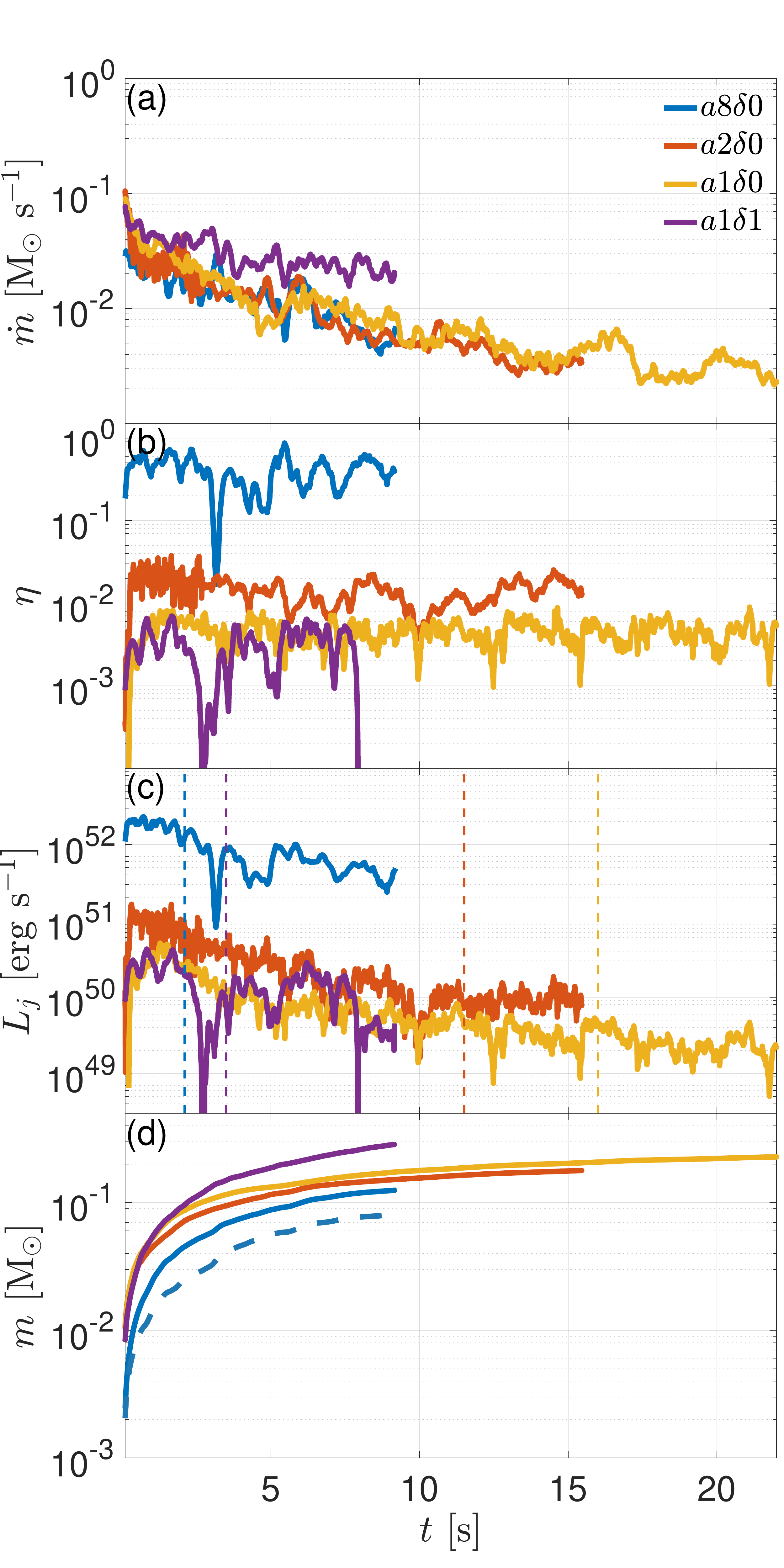}
        \caption{Smoothed time evolution of the numerical models, as measured at $ 3r_g $:
        \textbf{Panel (a):} The mass accretion rate onto the BH drops moderately over time, owing to suppression by the jet structure.
        \textbf{Panel (b):} The jet launching efficiency is in a good agreement with \citet{Lowell2023}, featuring $ \eta \simeq 0.5, 0.02, 0.005 $ for $ a = 0.8, 0.2, 0.1 $, respectively. \textbf{Panel (c):} Given the similar $ \dot{m} $ in all models, BHs with $ a = 0.1 $ power jets that are two orders of magnitude less luminous than jets from BHs with $ a = 0.8 $, having $ L_j \approx 10^{50}~\erg~\s^{-1}$ that is comparable with the characteristic GRB luminosity. The vertical dashed lines represent the breakout time of the outflow from the star, after which the luminosity can be observed.
        \textbf{Panel (d):} Accreted mass. The differences in the mass accretion rate (solid lines) lead to less total accreted mass onto the BH when the jet launching and propagation are more efficient. The high efficiency $ \eta \simeq 0.5 $ when $ a = 0.8 $ indicates that half of the accreted mass is expelled as outflows, resulting in BH mass gain that is only half of the total accreted mass (dashed line).
        }
        \label{fig:launching}
    \end{figure}

    Figure~\ref{fig:launching} depicts the time evolution of the BH accretion and jet launching quantities in the different models. Fig.~\ref{fig:launching}(a) features comparable accretion rates in all models, with a moderate decrease in $ \dot{m} $ over time, compared to the constant accretion rate expected from Eq. \eqref{eq:mdot}. The discrepancy originates in the suppression of the mass accretion rate by the laterally extended structure of the shocked material around the jet. The weak dependency of the lateral structure on the jet power explains the comparable accretion rates between the different models. However, in the presence of a low density region on the polar axis (model $ \C $), the jet propagates fast without spilling most of its energy to the jet backflows, so its effect on the mass accretion rate diminishes. {In particular, model $ \C $ features $ \dot{m} \propto t^{-1/4} $, whereas other models exhibit $ \dot{m} \propto t^{-3/4} $.}

    Fig.~\ref{fig:launching}(b) shows that the jet electromagnetic efficiency is dictated by the BH spin, and is consistent with Eq. \eqref{eq:eta}. The rapidly spinning BH (blue line) reaches efficiency of $ \eta \simeq 0.5 $, whereas slowly spinning BHs exhibit $ \eta \approx  10^{-2} $. All models reach those efficiencies early on, within $ 0.3 $ s after the BH formation, and remain quasi-steady. The jet power, as shown in Fig.~\ref{fig:launching}(c), is the product of the mass accretion rate and launching efficiency, and thus it is also solely governed by the BH spin. While rapidly rotating BHs ($ a = 0.8 $) produce jets with $ L_j \approx 10^{52}~\erg~\s^{-1} $ that are on the high end of the GRB energy distribution, slower ones with $ a = 0.2 $ give rise to jets with a typical GRB luminosity, $L_j \approx 10^{50}\erg~\s^{-1} $. This is the first time that a slowly spinning BH is shown to be able to launch a relativistic jet with such power into a collapsing stellar envelope. This result is consistent with the jet launching criterion that depends only on the magnetic field threading the BH, and with the prediction that a BH with $ a \approx 0.1 $ launches a jet with a typical GRB energy.
    {Due to the time evolution in the mass accretion rate, the jet luminosity also exhibits mild time evolution. Such mild evolution might be still consistent with observations since the jet is only observed after it breaks out from the star, between $ \sim 10 $ s and a few dozen seconds, namely the observed jet evolves over less than an order of magnitude in time, thus by merely a factor of a few in luminosity. Nonetheless, obtaining constant mass accretion and jet luminosity is possible with milder density profiles of $ \alpha = 1 $ \citep{Gottlieb2022a}, which are also roughly consistent with the models of \citet{Halevi2023}.}

    The roughly constant accretion rate leads to a linear growth in the accumulated accreted mass on the BH (Fig.~\ref{fig:launching}(d)). In model $ \C $, where the accreted mass is the highest (and the unbound mass is the lowest, see Tab.~\ref{tab:models}), the extrapolation of the observed linear growth suggests that the BH mass doubles after $ \approx 100~\s $. In model $ \Ah $, the efficiency of $ \eta \approx 0.5 $ implies that about half of the accreted mass energy is converted to jet power, so that the BH gains only half of the accreted mass (dashed line). The lower accreted mass in model $ \Ah $ results in a milder BH mass gain after its formation.
    This constitutes another argument against rapidly spinning BHs - the observed mass gap between NSs and BHs \citep[e.g.,][]{Ozel2010,Farr2011} indicates that after BHs form, they continue to accrete mass that is at least comparable to their natal mass during the stellar collapse \citep{Belczynski2012}. In order to self-consistently assess the long-term effects of initially rapidly spinning BHs, one needs to consider the BH spin-down, which we address in a companion paper \citep{Jacquemin2023}.

    \subsection{Jet propagation}\label{sec:propagation}

\begin{figure}
        \includegraphics[scale=0.106]{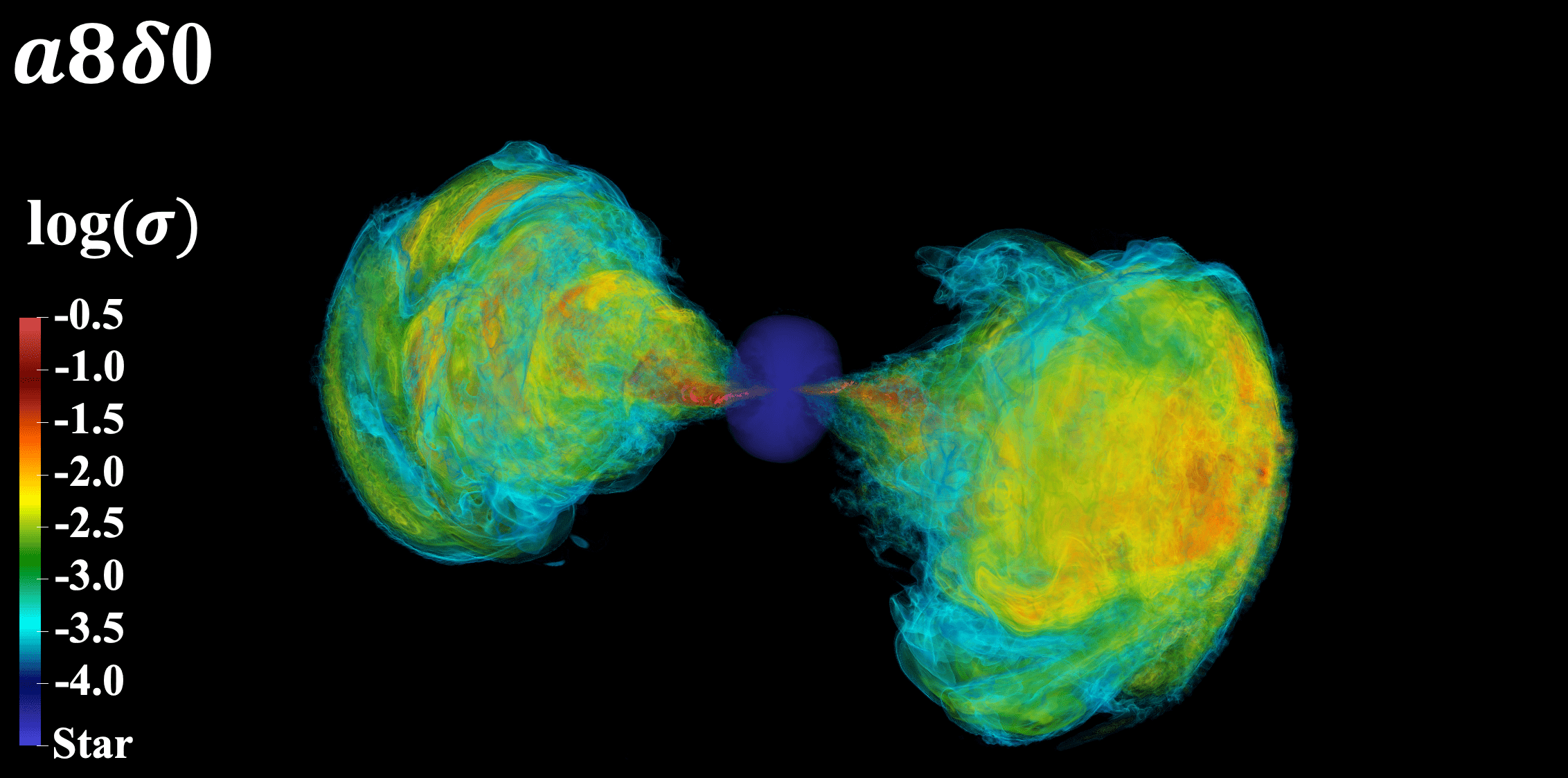}
        \includegraphics[scale=0.106]{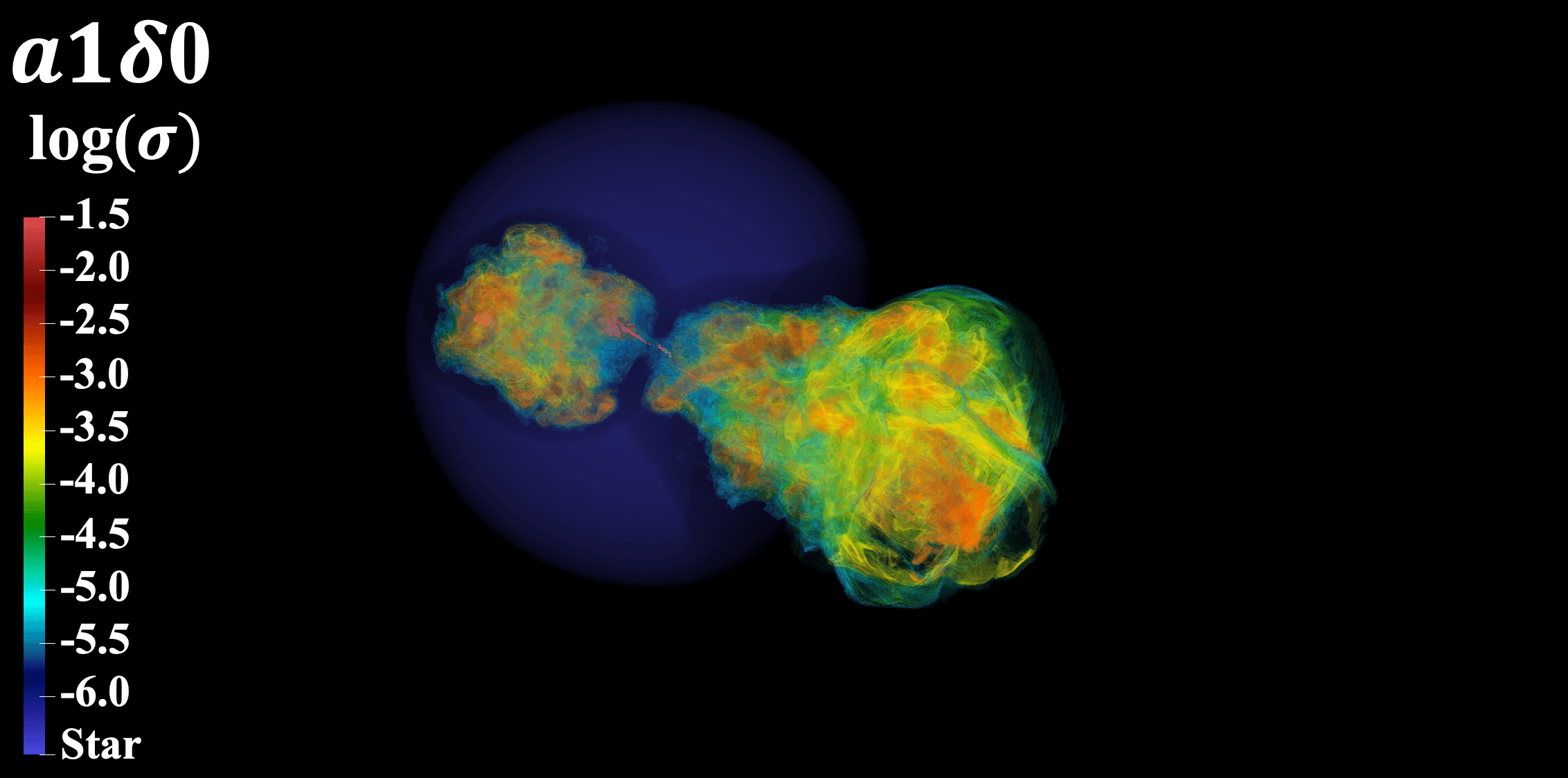}
        \includegraphics[scale=0.106]{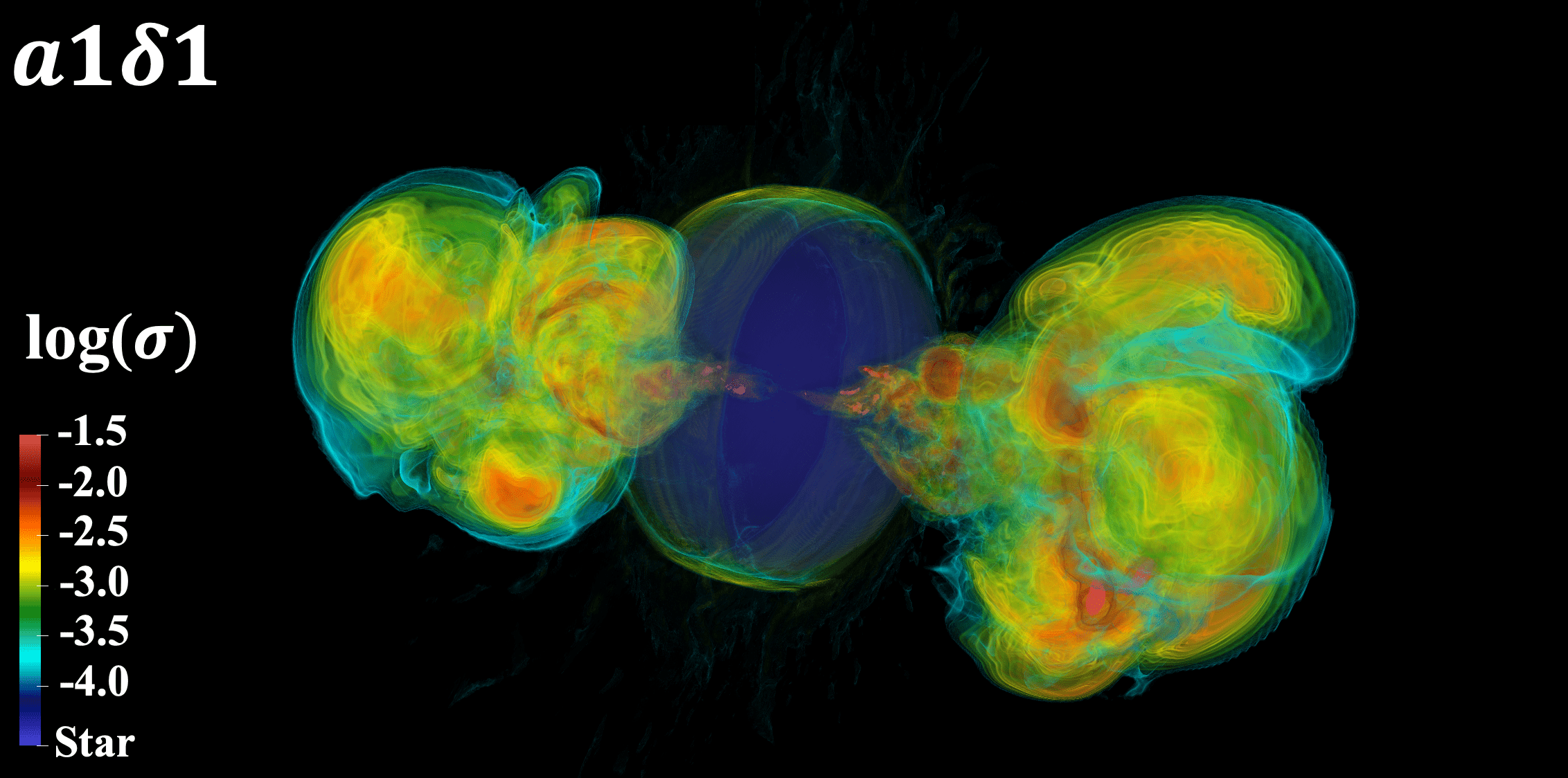}
        \caption{3D rendering of the jet magnetization after breakout in models $ \Ah $ (top), $ \Al $ (middle) and $ \C $ (bottom) at $ t = t_s $. While all jets find their way out from the star (dark blue), in model $ \Ah $ the jet retains a mild magnetization and features a stratified structure, whereas in models $ \Al $ and $ \C $ the jets fail to remain intact as they lose most of their energy to mixing with the star.
        }
        \label{fig:maps}
    \end{figure}

    We find that slowly spinning collapsar BHs power jets with $ L_j \approx 10^{50}~\erg~\s^{-1} $, which are favored by GRB observations. While hydrodynamic and weakly-magnetized jets of that power were previously shown to successfully break out from stellar envelopes \citep{Lopez-camara2013,Lopez-camara2016,Ito2015,Ito2019,Harrison2018,Gottlieb2019,Gottlieb2020b,Gottlieb2021a}, no first-principles numerical models exist for Poynting-flux dominated collapsar jets of that power. Here we find, for the first time, that when a jet with a typical GRB energy is launched into a spherical envelope (model $ \Al $), it breaks apart and invests almost all its energy in the expansion of a sub-relativistic ($ v \approx 0.1 $ c) shocked stellar material (see Figure~\ref{fig:maps}). As the spherical shock breaks out, it might power a low luminosity GRB or mildly relativistic transients such as fast blue optical transients \citep{Gottlieb2022b}. A detailed calculation of its electromagnetic signature will be conducted in future work.

    The disintegration of the jet might be attributed to its initial high magnetization, which in low power jets, gives rise to kink instabilities. However, the weak dependency of the jet kink instability criterion on the jet luminosity to density ratio, $ \sim (L/\rho)^{1/6} $, and the uncertainty of the stability critical value \citep{Bromberg2016}, make it difficult to determine whether the kink instability is responsible for the jet dissipation. An alternative explanation for the jet difficulty to efficiently pierce through the stellar envelope is its intermittency along the axis of propagation, which in lower power jets may lead to strong baryon contamination that destroys the jet \citep{Gottlieb2020a,Gottlieb2021b}. The intermittent jet structure emerges due to the abrupt nature of the central engine, and the jet wobbling motion, caused by the tilt of the disk \citep{Gottlieb2022c}, which launches the jet in different directions. In other words, the effective jet head cross section becomes too large, and considering the lower jet power, its luminosity density is too low to enable an efficient jet propagation through the star.

    Regardless of the physical mechanism responsible for the jet destruction, a lower mass density along the rotational axis of the star may mitigate the jet propagation. In our anisotropic model with $ \delta = 1 $, the polar ($ \theta \lesssim \theta_j $) isotropic equivalent mass is about 15 times lower than the isotropic case, equivalent to increasing the jet power by the same factor. Our simulation with $ \delta = 1 $ exhibits jets that quickly break out from the star, with an average head velocity inside the star of 0.4 c. Although the jet in an anisotropic star quickly breaches the envelope, its breakout characteristics are similar to those observed for jets in models $ a1\delta0, a2\delta0 $, and are inconsistent with those inferred from GRB observables. Fig.~\ref{fig:maps} portrays 3D renderings of the outflow magnetization in models $ \Ah $ (top), $ \Al $ (middle) and $ \C $ (bottom) after breaking out from the star (dark blue). In the first case, the powerful jet remains collimated and features a moderate level of magnetization. Both weaker jets dissipate most of their magnetic energy by mixing with the star, and break out with a negligible degree of magnetization while losing their collimated structure.

    \begin{figure}
        \includegraphics[scale=0.33]{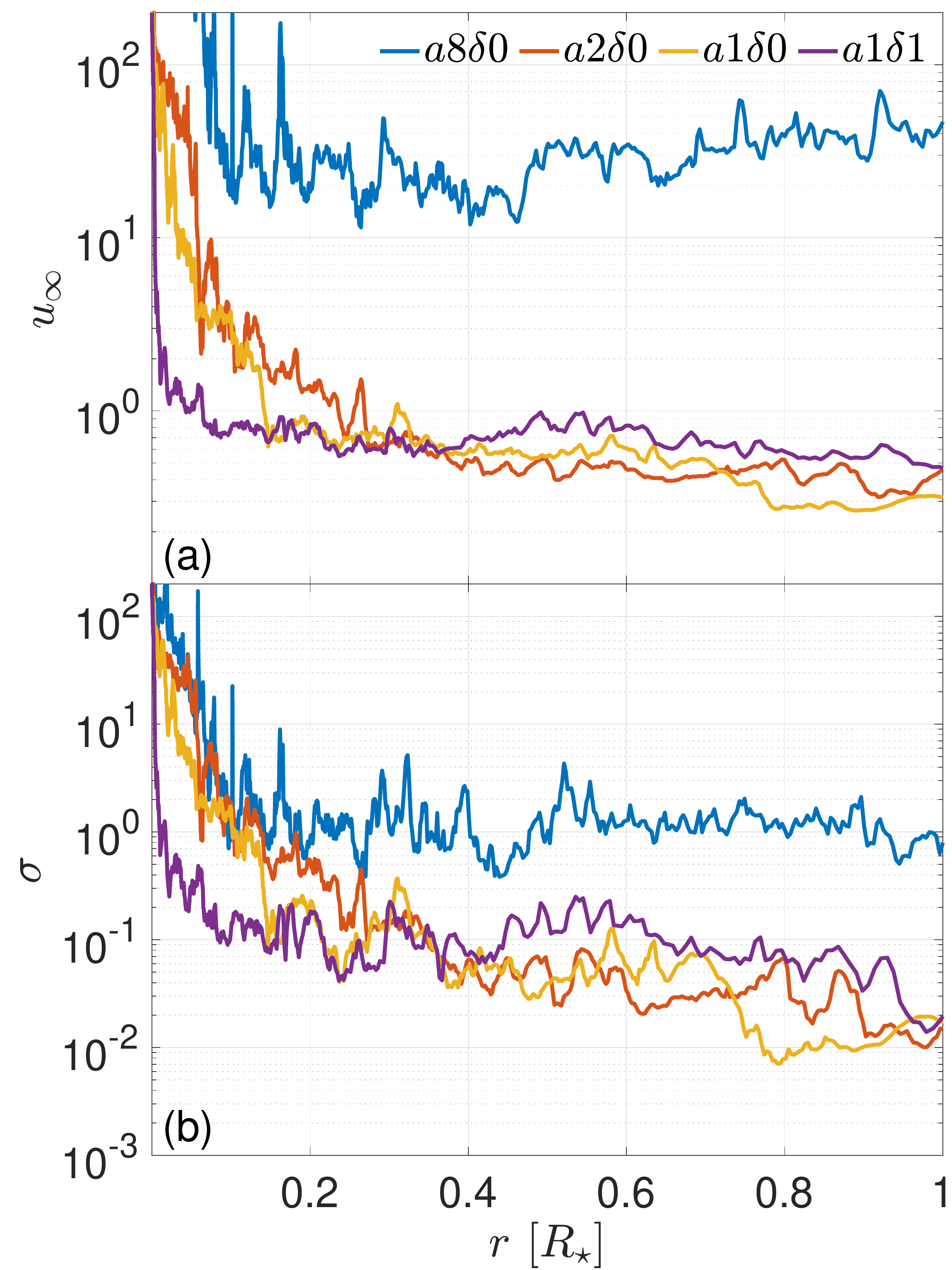}
        \caption{Radial profiles of the asymptotic proper-velocity \textbf{(a)} and magnetization \textbf{(b)} of the jets inside the star, calculated by the maximum value at each radius at time $ t_s $. The powerful jet (model $ \Ah $) maintains relativistic velocities and mild magnetization, whereas the weaker jets undergo strong mixing with the star so they exhibit a mildly-relativistic breakout with a low magnetization.
        }
        \label{fig:radial_profiles}
    \end{figure}
    
    Figure~\ref{fig:radial_profiles} depicts the radial profiles of the jets' asymptotic proper-velocity $ u_\infty $ (panel (a)) and magnetization $ \sigma $ (panel (b)), defined as
    \begin{equation}\label{eq:uinf}
        u_\infty \equiv \sqrt{u_t^2\left(1+\frac{4p}{\rho c^2}+\sigma\right)^2-1}~,
    \end{equation}
    where $ \rho $ is the comoving mass density, $ p $ is the comoving gas pressure, $ u_t $ is the covariant time component of the four-velocity, $ \sigma = \frac{B^2}{8\pi\rho c^2} $ is the magnetization, and $ B $ is the comoving magnetic field.
    The powerful jet launched by the rapidly spinning BH retains a moderate magnetization level and reaches ultra-relativistic velocities of GRBs (blue lines). Conversely, the jets powered by the slowly spinning BHs lose most of their magnetic energy deep inside the stellar core due to mixing. As a result, they lose their ability to convert that energy to kinetic form and reach relativistic velocities. We emphasize that although mixing reduces the jets velocity, their power upon breakout remains similar to their power upon launching. This raises the question of how such jets can reach asymptotic relativistic velocities to match both the GRB power and velocity. This problem might be alleviated if the jets were to be launched with $ \sigma_0 \gg 10^3 $, similar to the trend observed between different $ \sigma_0 $ values in \citet{Gottlieb2022c}. Unfortunately, present-day numerical codes cannot handle jet evolution at such high $ \sigma $ values.
    
    \section{Discussion}\label{sec:conclusions}
    In this \emph{Letter} we show that theoretical considerations combined with GRB observations support the idea that collapsar BHs are likely born slowly spinning. The reasoning relies on several straightforward arguments. The first is that the magnetic flux is saturated in a MAD state (if its value is below saturation, the jet fails to launch). Hence, the magnetic-driven jet power depends only on the mass accretion rate and jet launching efficiency. \citet{Halevi2023} recently showed that at the time of BH formation, the inner stellar envelope has a universal radial density profile with a power-law $ -1.5 $. This is translated to a roughly constant mass accretion rate, that for a typical progenitor star with a few tens of solar mass, is $ \dot{m} \simeq 10^{-2} \msun~\s^{-1} $. Numerical models also feature constant jet launching efficiencies, so a constant accretion rate also implies no time evolution in the GRB light curve, as suggested by observations. This leaves the jet power to depend solely on the value of the jet launching efficiency, which in turn depends only on the BH spin. GRB prompt emission observations thus provide a direct connection between observables and BH spin.
    
    We show that wobbling jets, as found in our simulations, require a dimensionless spin of $ a \lesssim 0.5 $ in order to match observed luminosities. If the jet is roughly axisymmetric, as traditional jet structure models suggest, a milder $ a \approx 0.2 $ is needed to launch a jet that produces the observed GRB luminosities. The above moderate spins correspond to low jet launching efficiency, implying that most of the accreted energy onto the BH is used for its mass growth rather than launching jets. In addition to the jet motion and opening angle, there are another two important caveats to this result: (i) the uncertainty in the jet radiative efficiency -- if $ \epsilon_\gamma \ll 1 $, then the inferred BH spin would be significantly larger; (ii) in our analysis of inferring the BH spin for axisymmetric jets, we assume that the entire jet is observed, thus ignore cases in which the jet is observed slightly off-axis such that only part of its energy reaches the observer \citep[e.g.,][]{Ito2019}. If most GRBs are observed off-axis, then the jet energy, and the BH spin, could be significantly higher.
    Regardless of the value of the inferred spin, this value might hold true for the entire core-collapse BH population, unless there is an anti-correlation between the magnetic field strength in the star and the angular momentum of the star (e.g., through the magnetic Tayler instability), in which case BHs without jets spin faster. Namely, GRB observations indicate that the natal spin of the majority of newly formed BHs is small, otherwise there will be an excess of very powerful GRB jets in the Universe.

    We verify our results by carrying out first-principles collapsar simulations, and show for the first time that slowly spinning BHs can launch relativistic jets with a typical GRB power, thereby supporting the above theoretical arguments. We find that the jet power does not change over time, but less powerful jets undergo intense mixing, even when a lower density region along the poles is present. Consequently, the jets escape from the star being mildly-relativistic and cannot reproduce the GRB observables.
    A possible solution to this problem is initial jet magnetization of $ \sigma_0 \gg 10^3 $, which may enable the jets to remain relativistic even after dissipating a substantial fraction of their magnetic energy.
        
    While a moderate BH spin of $ a \simeq 0.2 $ can generate typical GRB jets of power $ L_j \approx 10^{50}~\erg~\s^{-1} $, the GRB energy distribution spans a vast range of many orders of magnitude. At the low end of the GRB luminosity distribution lie jets with $ L_j \lesssim 10^{48}~\erg~\s^{-1} $\citep{Wanderman2010,Shahmoradi2015}, which begs the question of what factors would support their emergence. Regardless of how such jets find their way out of the collapsing star, the fixed jet power throughout its propagation and Eq.~\eqref{eq:eta} dictate that these jets must be launched from a BH with a spin of $ a \simeq 0.02 $. However, for such spin, the disk winds may outshine the jet and disrupt its emergence. Furthermore, in \citet{Jacquemin2023} we show that BHs with an initial spin of $ a \lesssim 0.1 $ inevitably spin up to $ a \simeq 0.1 $ before the jet breaks out. Therefore, low power jets likely emerge from a BH with $ a \approx 0.1 $, but with a lower accretion rate.
    
    At the high end of the GRB energy distribution, powerful jets require a BH spin close to unity. However, in \citet{Jacquemin2023} we show that the BH likely spins up to high spins rather than having a high natal spin. This implies that conceivably all collapsar BHs are born slowly spinning, and when the magnetic field profile of the star is such that the development of the MAD state is delayed, the BH may spin up to $ a \approx 0.5 $ before the MAD state fully develops. Once the system is MAD, the BH will spin down and reach a low final spin, $a \approx 0.1$. Thus, the high spin is only achieved for a relatively short time.

    As most collapsar BHs  have $ a \lesssim 0.2 $, our simulations indicate that they would produce jets that struggle to break out relativistically from stars, and it is likely that some of those jets would fail to generate the GRB emission. Instead, the jets will energize the expansion of the shocked jet material that will ultimately break out and radiate a softer emission that could be associated with sub- and mildly- relativistic transients. This conclusion is obtained for the first time from a computational perspective, supporting the idea that many collapsar jets are choked based on the GRB duration distribution \citep{Bromberg2012}.

    Here we study the effect of the initial BH spin on jet launching, assuming a BH spin that does not change in time. In reality, the BH spins up by accreting angular momentum from the infalling gas, and spins down by utilizing its rotational energy to launch the jets. These effects are particularly important when the BH spin is far from the equilibrium spin $ a \approx 0.1 $ \citep{Lowell2023}. In a companion paper \citep{Jacquemin2023}, we show that taking into account the BH spin evolution does not change our conclusion that disfavors rapidly spinning BHs. The reason is that the mass accretion rate is not high enough to introduce a significant spin-down within the jet breakout time from the star. Furthermore, in \citet{Jacquemin2023} we show that even if the BH were to be initially rapidly spinning, for any reasonable set of physical parameters, it eventually spins down close to the equilibrium spin.
                
	\begin{acknowledgements}

    We thank the referee for helpful comments.
    OG is supported by a CIERA Postdoctoral Fellowship.
    OG and AT acknowledge support by Fermi Cycle 14 Guest Investigator program 80NSSC22K0031.
    JJ and AT acknowledge support by the NSF AST-2009884 and NASA 80NSSC21K1746 grants.
    BL acknowledges support by a National Science Foundation Graduate Research Fellowship under Grant No. DGE-2234667. BL also acknowledges support by a Illinois Space Grant Consortium (ISGC) Graduate Fellowship supported by a National Aeronautics and Space Administration (NASA) grant awarded to the ISGC.
    AT was also supported by NSF grants
    AST-2107839, 
    AST-1815304, 
    AST-1911080, 
    AST-2206471,
    OAC-2031997, 
    and NASA grant 80NSSC18K0565. 
    Support for this work was also provided by the National Aeronautics and Space Administration through Chandra Award Number TM1-22005X issued by the Chandra X-ray Center, which is operated by the Smithsonian Astrophysical Observatory for and on behalf of the National Aeronautics Space Administration under contract NAS8-03060.
    This research used resources of the Oak Ridge Leadership Computing Facility, which is a DOE Office of Science User Facility supported under Contract DE-AC05-00OR22725. An award of computer time was provided by the ASCR Leadership Computing Challenge (ALCC), Innovative and Novel Computational Impact on Theory and Experiment (INCITE), and OLCF Director's Discretionary Allocation  programs under award PHY129. This research used resources of the National Energy Research Scientific Computing Center, a DOE Office of Science User Facility supported by the Office of Science of the U.S. Department of Energy under Contract No. DE-AC02-05CH11231 using NERSC award ALCC-ERCAP0022634.

	\end{acknowledgements}
	
	\section*{Data Availability}
	
	The data underlying this article will be shared upon reasonable request to the corresponding author.
	
	\bibliography{refs} 

\end{document}